# TESTING RELATIVISTIC GRAVITY AND DETECTING GRAVITATIONAL WAVES IN SPACE[*]

WEI-TOU NI

*Center for Gravitation and Cosmology, Purple Mountain Observatory,
Chinese Academy of Sciences, No. 2, Beijing W. Rd., Nanjing, 210008 China
National Astronomical Observatories, Chinese Academy of Sciences,
Beijing, 100012 China*

*e-mail: wtni@pmo.ac.cn*

For testing gravity and detecting gravitational waves in space, deep-space laser ranging using drag-free spacecraft is a common method. Deep space provides a large arena and a long integration time. Laser technology provides measurement sensitivity, while drag-free technology ensures that gravitational phenomenon to be measured with least spurious noises. In this talk, we give an overview of motivations and methods of various space missions/proposals testing relativistic gravity and detecting gravitational waves, and refer to various references.

## 1. Introduction - Mapping the Gravitational Field in the Solar System

The solar-system gravitational field is determined by three factors: the dynamic distribution of matter in the solar system; the dynamic distribution of matter outside the solar system (galactic, cosmological, etc.) and gravitational waves propagating through the solar system. Different relativistic theories of gravity make different predictions of the solar-system gravitational field. Hence, precise measurements of the solar-system gravitational field test these relativistic theories, in addition to enabling gravitational wave observations, determination of the matter distribution in the solar-system and determination of the observable (testable) influence of our galaxy and cosmos. The tests and observations from various missions with various configurations could include: (i) a precise determination of the relativistic parameters with 3-5 orders of magnitude improvement over previous measurements; (ii) a 1-2 order of magnitude improvement in the measurement of $\dot{G}$; (iii) a precise determination of any anomalous, constant acceleration $A_a$ directed towards the Sun; (iv) a measurement of solar angular momentum via the Lense-Thirring effect; (v) the detection of solar g-mode oscillations via their changing gravity field, thus, providing a new eye to see inside the Sun; (vi) precise determination of the planetary orbit elements and masses; (vii) better determination of the orbits and masses of major asteroids; (viii) exploring the dynamic distribution of matter outside the solar system (galactic, cosmological, etc.) and testing large scale gravitational theories; (ix) detection and observation of gravitational waves from massive black holes and galactic compact binary stars in the low and middle frequency range 100 nHz to 10 Hz; and (x) exploring background gravitational-waves [1].

---

[*]This work is supported by the National Natural Science Foundation of China (Grant Nos. 10778710 and 10875171).

## 2. Methods of Measurement

For testing gravity and detecting gravitational waves in space, deep-space laser ranging using drag-free spacecraft is a common method. Deep space provides a large arena and a long integration time. Laser technology provides measurement sensitivity, while drag-free technology ensures that gravitational phenomenon to be measured with least spurious noises. Up to now, microwave technology is used for measuring the solar-system and testing solar-system gravity. However, this will be replaced by more precise laser technology. The precision has already been demonstrated by lunar laser ranging [2]. For last forty years, we have seen great advances in the dynamical testing of relativistic gravity. This is largely due to interplanetary radio ranging and lunar laser ranging [3]. Interplanetary radio ranging and tracking provided more stimuli and progresses at first. However with improved accuracy of 1 mm from 20-30 cm and long-accumulation of observation data, lunar laser ranging reaches similar accuracy in determining relativistic parameters as compared to interplanetary radio ranging.

There are various interplanetary optical missions proposed. As an example of interplanetary missions, we mention ASTROD; the baseline scheme of ASTROD is to have two spacecraft in separate solar orbits and one spacecraft near the Earth-Sun L1/L2 point carrying a payload of a proof mass, two telescopes, two 1-2 W lasers with spares, a clock and a drag-free system ranging coherently among one another using lasers. [1]

For technology development, we refer the readers to LISA proceedings and ASTROD proceedings.

## 3. Demonstration of Interplanetary Laser Ranging

Interplanetary laser ranging has been demonstrated by MESSENGER (MErcury Surface, Space ENvironment, GEochemistry, and Ranging) [4-6]. The MESSENGER spacecraft, launched on 3 August 2004, is carrying the Mercury

Laser Altimeter (MLA) as part of its instrument suite on its 6.6-year voyage to Mercury. Between 24 May, 2005 and 31 May, 2005 in an experiment performed at about 24 million km before an Earth flyby, the MLA on board MESSENGER spacecraft performed a raster scan of Earth by firing its Q-switched Nd:YAG laser at an 8 Hz rate. Pulses were successfully received by the 1.2 m telescope aimed at the MESSENGER spacecraft in the NASA Goddard Geophysical and Astronomical Observatory at Gaddard Space Flight Center (GSFC) when the MLA raster scan passed over the Earth station. Simultaneously, a ground based Q-switched Nd:YAG laser at GSFC's 1.2 m telescope was aimed at the MESSENGER spacecraft. Pulses were successfully exchanged between the two terminals. From this two-way laser link, the range as a function of time at the spacecraft over $2.39 \times 10^{10}$ m (~ 0.16 AU) was determined to ± 0.2 m (± 670 ps): a fractional accuracy of better than $1 \times 10^{-11}$.

A similar experiment was conducted by the same team to the Mars Orbiter Laser Altimeter (MOLA) on board the Mars Global Surveyor (MGS) spacecraft in orbit about Mars [6]. At that time, the MOLA laser was no longer operable after a successful topographic mapping mission at Mars. The experiment was one way (uplink) and the MOLA detector saw hundreds of pulses from 8.4 W Q-switched Nd:YAG laser at GSFC.

## 4. Missions for Testing Relativistic Gravity

In this section, we briefly review various missions for testing relativistic gravity and illustrate by looking into various ongoing / proposed experiments related to the determination of the PPN space curvature parameter γ. Some motivations for determining γ precisely to $10^{-5} - 10^{-9}$ are given in [7, 8].

First, we mention two recently completed experiments --- Cassini experiment and GP-B experiment. Cassini experiment is the most precise experiment measuring the PPN space curvature parameter γ up to now. In 2003, Bertotti, Iess and Tortora [9] reported a measurement of the frequency shift of radio photons due to relativistic Shapiro time-delay effect from the Cassini spacecraft as they passed near the Sun during the June 2002 solar conjunction. From this measurement, they determined γ to be 1.000021 ± 0.000023. GP-B experiment used quartz gyro at low-temperature to measure the Lense-Thirring precession and the geodetic precession in a polar orbit of the earth. The geodetic precession gives a measure of γ. The precision from their current analysis is around 0.003 [10].

Bepi-Colombo [11] is planned for a launch in 2013 to Mercury. A simulation predicts that the determination of γ can reach $2 \times 10^{-6}$ [12].

GAIA (Global Astrometric Interferometer for Astrophysics) [13] is an astrometric mission concept aiming at the broadest possible astrophysical exploitation of optical interferometry using a modest baseline length (~3m). GAIA is planned to be launched in 2013. At the present study, GAIA aims at limit magnitude 21, with survey completeness to visual magnitude 19-20, and proposes to measure the angular positions of 35 million objects (to visual magnitude V=15) to 10 μas accuracy and those of 1.3 billion objects (to V=20) to 0.2 mas accuracy. The observing accuracy of V=10 objects is aimed at 4 μas. To increase the weight of measuring the relativistic light deflection parameter γ, GAIA is planned to do measurements at elongations greater than 35° (as compared to essentially 47° for Hipparcos) from the Sun. With all these, a simulation shows that GAIA could measure γ to $1 \times 10^{-5} - 2 \times 10^{-7}$ accuracy [14].

ASTROD I is a first step towards ASTROD. Its scheme is to have one spacecraft in a Venus-gravity-assisted solar orbit, ranging optically with ground stations with less ambitious, but still significant scientific goals in testing relativistic gravity. In the ranging experiments, the retardations (Shapiro time delays) of the electromagnetic waves are measured to give γ. In the astrometric experiments, the deflections of the electromagnetic waves are measured to give γ. These two kinds of experiments complement each other in determining γ. The Cosmic Vision ASTROD I (Single Spacecraft Astrodynamical Space Test of Relativity using Optical Devices) mission concept [15] is to use a drag-free spacecraft orbiting around the Sun using 2-way (both uplink and downlink) laser pulse ranging between Earth and spacecraft to measure γ and other relativistic parameters precisely. The γ parameter can be separated from the study of the Shapiro delay variation. The uncertainty on the Shapiro delay measurement depends on the uncertainties introduced by the atmosphere, timing systems of the ground and space segments, and the drag-free noise. A simulation shows that an uncertainty of $3 \times 10^{-8}$ on the determination of γ is achievable.

LATOR (Laser Astrometric Test Of Relativity) [16] proposed to use laser interferometry between two micro-spacecraft in solar orbits, and a 100 m baseline multi-channel stellar optical interferometer placed on the ISS (International Space Station) to do spacecraft astrometry for a precise measurement of γ.

For ASTROD (Astrodynamical Space Test of Relativity) [1], 3 spacecraft, advanced drag-free systems, and mature laser interferometric ranging will be used and the resolution is subwavelength. The accuracy of measuring γ and other parameters will depend on the stability of the lasers and/or clocks. An uncertainty of $1 \times 10^{-9}$ on the determination of γ is achievable in the time frame of 2025-2030.

Space mission proposals to the outer solar system for testing relativistic gravity and cosmology theories include Super-ASTROD [17], SAGAS [18] and Odyssey [19]. We refer the readers to the refereces mentioned.

## 5. Missions for Detecting Gravitational Waves

A complete classification of gravitational waves according to their frequencies is: (i) Ultra high frequency band (above 1 THz); (ii) Very high frequency band (100 kHz – 1 THz); (iii) High frequency band (10 Hz – 100 kHz); (iv) Middle frequency band ( 0.1 Hz – 10 Hz); (v) Low frequency band (100 nHz – 0.1 Hz); (vi) Very low frequency band (300 pHz – 100 nHz); (vii) Ultra low frequency band (10 fHz – 300 pHz); (viii) Hubble (extremely low) frequency band (1 aHz – 10 fHz); (ix) Infra-Hubble frequency band (below 1 aHz). The aims of gravitational-wave space missions are for detection of gravitational waves in the low frequency band (LISA, ASTROD, ASTROD-GW, and Super-ASTROD) and middle frequency band (DECIGO and Big Bang Observer). The space detectors are complimentary to the ground detectors which aim at detection of gravitational waves in the high frequency band.

LISA has nearly equilateral triangular spacecraft formation of armlength $5 \times 10^6$ km in orbit 20° behind earth.

ASTROD has a large variation in its triangular formation with armlength varying up to about 2 AU.

ASTROD-GW (ASTROD [Astrodynamical Space Test of Relativity using Optical Devices] optimized for Gravitation Wave detection) is an optimization of ASTROD to focus on the goal of detection of gravitational waves. The detection sensitivity is shifted 52 times toward larger wavelength compared to that of LISA. The scientific aim is focused for gravitational wave detection at low frequency. The mission orbits of the 3 spacecraft forming a nearly equilateral triangular array are chosen to be near the Sun-Earth Lagrange points L3, L4 and L5. The 3 spacecraft range interferometrically with one another with arm length about 260 million kilometers. After mission-orbit optimization, the changes of arm length are less than 0.0003 AU or, fractionally, less than $\pm 10^{-4}$ in ten years, and the Doppler velocities for the three spacecraft are less than ±4m/s. Both fit the LISA requirement and a number of technologies developed by LISA could be applied to ASTROD-GW. For the purpose of primordial GW detection, a 6-S/C formation for ASTROD-GW will be used for correlated detection of stochastic GWs.

The science goals for these missions are detection of Gravitational Waves (GWs) from (i) Supermassive Black Holes; (ii) Intermediate-Mass Black Holes;

(iii) Extreme-Mass-Ratio Black Hole Inspirals; (iv) Galactic Compact Binaries; and (v) Primordial Gravitational Wave Sources, Strings, Boson Stars etc.

For direct detection of primordial (inflationary, relic) GWs in space, one may go to frequencies lower or higher than the LISA [20] bandwidth, where there are potentially less foreground astrophysical sources to mask detection. DECIGO [21] and Big Bang Observer [22-23] look for GWs in the higher frequency range while ASTROD [1], ASTROD-GW [24] and Super-ASTROD [17] look for GWs in the lower frequency range. In the following section, we address to the issue of detectability of primordial gravitational waves.

## 6. The Detectability of Primordial Gravitational Waves

Inflationary cosmology is successful in explaining a number of outstanding cosmological issues including the flatness, the horizon and the relic issues. More spectacular is the experimental confirmation of the structure as arose from the inflationary quantum fluctuations. However, the physics in the inflationary era is unclear. Polarization observations of Cosmic Microwave Background (CMB) missions may detect the tensor mode effects of inflationary gravitational waves (GWs) and give an energy scale of inflation. To probe the inflationary physics, direct observation of gravitational waves generated in the inflationary era is needed. The following figure shows the sen that the direct observation of these GWs with sensitivity down to $\Omega_{gw} \sim 10^{-20}$-$10^{-23}$ is possible using present projected technology development if foreground could be separated (Figure 1).

## 7. Outlook

Space missions using optical devices will be important in testing relativistic gravity, measuring solar-system parameters and detecting gravitational waves. Laser Astrodynamics in the solar system envisages ultra-precision tests of relativistic gravity, provision of a new eye to see into the solar interior, precise measurement of $\dot{G}$, monitoring the solar-system mass loss, and detection of low-frequency gravitational waves to probe the early Universe and study strong-field black hole physics together with astrophysics of binaries. One spacecraft and multi-spacecraft mission concepts are in line for mission opportunities. In view of their importance both in fundamentals and in technology developments, mission concepts of this kind will be a focus in the near future.

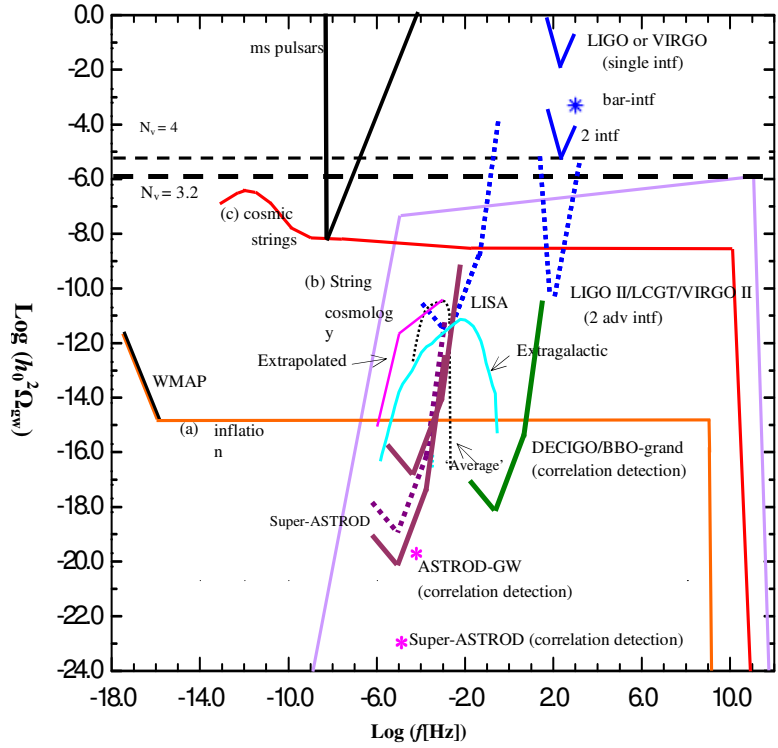

Figure 1. The stochastic backgrounds with bounds and GW detector sensitivities. (Adapted from Figure 3 and Figure 4 of Ref's 17 and 25 with sensitivity curve of DECIGO/BBO-grand added; the extragalactic foreground and the extrapolated foreground curves are from Ref. 27; see Ref's 17, and 25-28 for explanations).